\documentclass[preprint,12pt]{elsarticle}
\usepackage{xcolor}

\usepackage{amssymb}
\usepackage{amsmath}
\usepackage{tabularx}
\usepackage{booktabs}
\usepackage{tabularx}
\usepackage{svg}
\usepackage{multirow}
\journal{Nuclear Physics B}

\begin{document}

\begin{frontmatter}



\title{XT-REM: A Two-Component Model for Meta-Analysis of Extreme Event Proportions}


\author[label1]{Jovana Dedeić}
\author[label1]{Jelena Ivetić} 
\author[label1]{Srđan Milićević}
\author[label1]{Katarina Vidojević}
\author[label1]{Marija Delić}

\affiliation[label1]
{organization={University of Novi Sad, Faculty of Technical Sciences},
            addressline={Trg Dositeja Obradovića 6}, 
            city={Novi Sad},
            postcode={21000}, 
            country={Serbia}}

\begin{abstract}

In this paper, we introduce a novel model for the meta-analysis of proportions that integrates the standard random-effects model (REM) with an extreme value theory (EVT)–based component. The proposed model, named XT-REM (Extreme-Tail Random Effects Model), extends the classical REM framework by explicitly accounting for extreme proportions through a partial segmentation of the study set based on a predefined threshold. While the majority of proportions are modeled using REM, proportions exceeding the threshold are analyzed using the Generalized Pareto Distribution (GPD).

This formulation enables a dual interpretation of meta-analytic results, providing both an aggregate estimate for the central bulk of studies and a separate characterization of tail behavior. The XT-REM framework accommodates heteroskedastic variance structures inherent to proportion data, while preserving identifiability and consistency.

Using real-world data on immunotherapy-related adverse events, together with simulation studies calibrated to empirical settings, we demonstrate that XT-REM yields a comparable central estimate  while enabling a more explicit
assessment of tail behavior, including high-percentile extreme proportions. Compared with the classical REM, XT-REM achieves higher log-likelihood values and lower AIC, in the considered scenarios, indicating a better fit
within this modeling framework.

In summary, XT-REM offers a theoretically grounded and practically useful extension of random-effects meta-analysis, with potential relevance to clinical contexts in which extreme event rates carry important implications for risk assessment.
\end{abstract}




\end{frontmatter}
\section{Introduction}\label{intro}

In contemporary biomedical research, rare events with substantial clinical implications are increasingly being analyzed. One prominent example involves adverse events during immunotherapy, where event rates are typically low yet critically important for patient safety. Meta-analysis plays a key role in synthesizing information across multiple studies, particularly when individual studies have limited statistical power \cite{Borenstein2009, higgins2011}. Consequently, its application is far-reaching, spanning a wide range of scientific disciplines, as demonstrated by numerous studies \cite{ivetic2025,Schmid1998, Ke2015}.

Within standard meta-analytic frameworks, random-effects models (REM) are widely used. These models account for between-study heterogeneity, i.e., variability not attributable to sampling error alone. 
The estimation of between-study variance 
is a critical aspect of REM, with several estimators available, each with different properties in terms of bias and efficiency \cite{Veroniki2016}.
On the other hand, extreme value theory (EVT) is used to model tail behavior of distributions. Namely, the highest or lowest observed values. This approach is particularly valuable when analyzing rare but potentially serious outcomes, such as high rates of toxicity or severe complications \cite{Coles2001, Beirlant2006}. 
Indeed, EVT has already been applied in public‑health settings to estimate risks of rare but extreme events, e.g., for mortality spikes or surges in hospital admissions \cite{Thomas2016}.
The statistical literature has developed EVT-based models that incorporate regression components to estimate risk as a function of covariates, opening the possibility of linking EVT methods with hierarchical structures such as those found in meta-analysis, where between-study variability is naturally represented through random effects \cite{Coles1996}. 

Existing approaches to the meta-analysis of proportions in rare-event settings typically treat REM and EVT separately. REM models estimate central tendency and between-study variability \cite{Borenstein2009,higgins2009}, while EVT methods focus on modeling distribution tails and quantifying extreme values \cite{Coles2001, Beirlant2006}. However, neither approach alone enables simultaneous modeling of both central and extreme behavior within the hierarchical structure characteristic of meta-analysis.

Several statistical frameworks have been developed to combine bulk and tail modeling. Finite mixture models represent the overall distribution as a probabilistic combination of component distributions, allowing observations to arise from latent subpopulations \cite{Teicher1963, McLachlanPeel2000}. Alternatively, spliced (or composite) models combine distinct parametric families below and above a deterministic threshold, typically yielding a unified density over the entire support \cite{cooray2005}. These approaches provide flexible mechanisms for capturing heterogeneity and heavy-tailed behavior, but they differ in structure and inferential objectives. Within hierarchical and regression settings, recent studies have explored the integration of mixed effects models with extreme value theory to improve estimation for subgroups with small sample sizes \cite{Momoki2025}. This approach borrows strength across subgroups, reducing bias and variance in extreme value estimates. It is conceptually aligned with our aim to combine information across studies while modeling extreme outcomes,  yet it focuses on improving extreme-value estimation itself.

Although EVT models with regression components have been developed to allow extreme values to depend on covariates \cite{Coles1996}, and various nonlinear extensions of REM have been explored for rare-event settings \cite{kuss2015statistical, ibrahim2001bayesian}, a direct integration of EVT into the REM framework has not been systematically investigated. This gap motivates the development of a combined model capable of simultaneously:
\begin{itemize}
    \item estimating the global (aggregated) proportion of adverse events,
    \item identifying and quantifying extreme proportions that deviate from central tendencies,
    \item linking tail behavior to the latent (random-effects) study structure.
\end{itemize}

The meta-analysis of proportions presents specific methodological challenges beyond those encountered in effect-size-based syntheses. In clinical settings, such as the evaluation of immunotherapy-related adverse events, observed proportions can vary widely across studies, and extreme values often carry critical clinical information. Traditional REM provides a framework for estimating between-study heterogeneity, but it relies on assumptions that may not fully capture the specific characteristics of proportion data.
In meta-analyses of event proportions, variability arises at two levels. The within-study variance follows from the binomial sampling process and depends on the observed proportion and the study size, so precision differs across studies. The between-study variance, on the other hand, is introduced as a model parameter to capture heterogeneity under the Gaussian random-effects framework. When proportions are close to the boundaries, this variability can interact with the normal approximation in ways that have been discussed in the variance modeling literature \cite{CarrollRuppert1988, DavidianCarroll1987,hamza2008}.

{\it Research Objectives.}
The aim of this research is to develop a novel combined statistical model, named XT-REM (Extreme-Tail Random Effects Model), which integrates REM with EVT within a unified framework for estimating proportions in meta-analytic applications. Unlike probabilistic mixture models, which represent the overall distribution as a weighted combination of latent components, or spliced density models, which construct a single composite distribution across a threshold, XT-REM employs deterministic segmentation within a hierarchical meta-analytic structure. Its objective is not to define a unified density over the entire support, but rather to achieve inferential separation between central random-effects behavior and extreme tail dynamics. This separation preserves robust estimation of the central effect while explicitly modeling tail risk, making the framework particularly well suited for rare-event analyses in which both central and tail behavior must be quantified.

The specific objectives of the study are defined in the context of applications where accurate characterization of tail behavior is as important as estimating the central tendency:
\begin{itemize}
    \item to formalize the mathematical framework of the XT-REM model, which uses a REM component to describe the baseline distribution of study-level proportions and an EVT component for modeling extreme values,
    \item to theoretically investigate the identifiability and consistency of the model’s estimators, as well as their sensitivity to the threshold separating the central and tail regions,
    \item to conduct an empirical evaluation of XT-REM through simulation experiments with controlled parameters, and
    \item to apply the model to real-world data on immunotherapy-related adverse events, demonstrating its utility in a biomedical context.
\end{itemize}


{\it Structure of the paper.} 
The paper is organized as follows. In Section \ref{preliminaries}, we present the statistical foundations of the model, beginning with the classical REM framework and the key principles of EVT, to gradually introduce the proposed combined XT-REM model. This is followed by a formal definition of the model, including its component structure, the threshold separating central and extreme values, and the mechanism that links the two components. Particular attention is given to the theoretical properties of XT-REM, including considerations of identifiability, estimator consistency, and the implications of the modeling assumptions.
In Section \ref{novel_modeling} we describe the parameter estimation methodology, with an emphasis on maximum likelihood (ML) estimation and numerical aspects of the implementation. 
In Section~\ref{simulation}, we report the results of a simulation study designed to assess the performance of the proposed model under controlled conditions. Section~\ref{real_data_study} presents an application of XT-REM to real-world data on immunotherapy-related adverse events, illustrating its practical relevance and interpretability.
Finally, Section~\ref{conclusion} concludes the paper and outlines potential directions for future research.

\section{Preliminaries}\label{preliminaries}

To formally define the structure of our approach, we begin by revisiting the two well-known foundational components of the XT-REM model. The first is the classical REM, which captures between-study heterogeneity in the logit-transformed proportions. The second is the Generalized Pareto distribution from EVT, which models the behavior of proportions that exceed a predefined threshold.

\subsection{Random-Effects Model (REM)}\label{subsec:rem}

In the meta-analysis of proportions, let $r_i$ denote the number of subjects experiencing the event of interest in study $i$, and $n_i$ the total number of subjects. The observed proportion in study $i$ is defined as
\begin{equation}
p_i = \frac{r_i}{n_i}, \qquad i = 1,2, \dots, N,
\end{equation}
where $N$ denotes the total number of studies.

To improve statistical stability and obtain an approximately normal sampling distribution, the logit transformation is applied:
\begin{equation}\label{theta}
\theta_i = \ln{\left( \frac{p_i}{1 - p_i} \right)}.
\end{equation}

The logit transformation maps the unit interval $(0,1)$ to the real line $(-\infty,\infty)$ and is a standard link function in the meta-analysis of proportions \citep{Borenstein2009, Schwarzer2015}. Although the approximation may become less accurate for extremely small or large proportions, it performs well for the central range of values and is therefore widely used in practice.

Under the random-effects framework, the transformed proportion $\theta_i$ is assumed to follow the additive model
\begin{equation}\label{theta_i}
\theta_i = \mu + u_i + \varepsilon_i,
\end{equation}
where $\mu$ represents the overall mean logit-proportion, $u_i \sim N(0,\tau^2)$ denotes the study-specific random effect capturing between-study heterogeneity, and $\varepsilon_i \sim N(0,\sigma_i^2)$ represents the within-study sampling error.

Assuming independence of the random components, the total variance of the transformed outcome is
\begin{equation}
\operatorname{Var}(\theta_i) = \tau^2 + \sigma_i^2,
\end{equation}
where $\tau^2$ quantifies the between-study heterogeneity and $\sigma_i^2$ denotes the within-study variance on the logit scale.

Under a binomial sampling model, the variance of the empirical proportion is
\begin{equation}
\operatorname{Var}(p_i) = \frac{p_i(1-p_i)}{n_i}.
\end{equation}

Applying the delta method to the logit transformation yields the approximate variance on the logit scale
\begin{equation}
\sigma_i^2 \approx \frac{1}{n_i p_i(1-p_i)}.
\end{equation}

In practice, a commonly used plug-in estimator of the within-study variance is
\begin{equation}
\hat{\sigma}_i^2 = \frac{1}{r_i} + \frac{1}{n_i - r_i},
\end{equation}
which is algebraically equivalent to the delta-method approximation. This estimator becomes undefined when $r_i=0$ or $r_i=n_i$ (zero or all events). In such cases, a continuity correction (e.g., adding $0.5$ to both $r_i$ and $n_i-r_i$) is typically applied.

In the random-effects model, studies are weighted according to the inverse of their total variance,
\begin{equation}
w_i = \frac{1}{\tau^2 + \sigma_i^2}.
\end{equation}
This inverse-variance weighting arises naturally from maximum likelihood estimation and ensures that studies with higher precision contribute more to the pooled estimate, while still accounting for between-study heterogeneity.

In the proposed XT-REM framework, the logit-normal random-effects model is used primarily to describe the central portion of the data, while extreme proportions, where the normal approximation becomes unreliable, are modeled separately using extreme value theory.

An alternative approach to addressing heavy-tailed study distributions in meta-analysis involves the use of Student’s t-distribution for random effects instead of the Gaussian distribution. While the t-distribution ensures robust inference by symmetrically thickening the tails, the XT-REM framework proposed in this study addresses this challenge through a more flexible semi-parametric solution. By explicitly employing the GPD for studies exceeding a predefined threshold, XT-REM captures asymmetric extreme behavior and provides a distinct characterization of the tail-features that are not attainable through simple distributional substitution, such as the Student’s t-distribution, as considered in flexible and robust meta-analytic approaches \cite{noma2022meta, panagiotopoulou2025}. This distinction is particularly important in applications in which extreme values differ structurally from the bulk of the data.

\subsection{Extreme Value Theory}
\label{subsec:evt}

Extreme Value Theory deals with modeling the behavior of distribution tails, i.e., values that significantly deviate from the usual range of observed data.

In the context of meta-analysis of proportions, EVT can be used to model those studies whose proportions significantly exceed (or fall below) a certain threshold \(u\). In this work, we focus on modeling the right tail of the distribution, as high proportions are particularly important in the application to adverse events, representing a potentially elevated risk. In the case of interest in low values (left tail), the analytical procedure would be carried out symmetrically, after transforming the data in the negative direction.

For the observed proportions \(p_i\) that exceed the threshold \(u\), i.e., \(p_i > u\), the GPD is used as a model for the excess over the threshold:
\begin{equation}\label{Y}
Y_i = p_i - u \quad \text{such that} \quad Y_i > 0.
\end{equation}
The GPD has the following probability density function (PDF):
\begin{equation}
f_{\xi,\beta}(y) =
\begin{cases}
\displaystyle \frac{1}{\beta}\left(1 + \frac{\xi y}{\beta}\right)^{-\frac{1}{\xi}-1}, & \xi \neq 0, \\[12pt]
\displaystyle \frac{1}{\beta} \exp\left(-\frac{y}{\beta}\right), & \xi = 0,
\end{cases}
\quad y > 0,\; \beta > 0,
\label{eq:gpd_pdf}
\end{equation}

\noindent where:
\begin{itemize}
    \item \(\xi\) is the shape parameter, which determines the heaviness of the tail:
    \begin{itemize}
        \item \(\xi > 0\): heavy tails (Pareto Type I),
        \item \(\xi = 0\): exponential distribution (limiting case),
        \item \(\xi < 0\): lighter tails with a finite upper bound.
    \end{itemize}
    \item \(\beta\) is the scale parameter, which stretches or contracts the distribution.
\end{itemize}





The fundamental differences and complementary roles of the REM and GPD components within the proposed framework are summarized in Table \ref{tab:rem_vs_gpd}.
\begin{table}[ht]
\centering
\caption{Comparison of Model Components within the XT-REM Framework}
\label{tab:rem_vs_gpd}
\resizebox{\textwidth}{!}{
\begin{tabular}{lll}
\toprule
\textbf{Feature} & \textbf{REM} & \textbf{GPD} \\
\midrule
\textbf{Data Scope} & Central bulk ($p_i \leq u$) & Right tail ($p_i > u$) \\
\textbf{Transformation} & Logit: $\ln\!\left(\frac{p}{1-p}\right)$ & Excess: $Y = p - u$ \\
\textbf{Distribution} & Normal: $\mathcal{N}(\mu, \tau^2 + \sigma_i^2)$ & GPD: $f_{\xi,\beta}(y)$ \\
\textbf{Key Parameters} & Mean ($\mu$), Heterogeneity ($\tau^2$) & Shape ($\xi$), Scale ($\beta$) \\
\textbf{Tail Behavior} & Exponential decay (thin tail) & Polynomial / heavy / finite tail ($\xi$) \\
\textbf{Primary Role} & Estimating average effect size & Quantifying extreme risk and tail behavior \\
\bottomrule
\end{tabular}
}
\end{table}

In the next section, we integrate the GPD into the hierarchical REM framework. Specifically, we allow the tail parameters to vary with study-specific random effects, providing a consistent model for both the central bulk and extreme proportions.

\section{Novel Modeling Framework XT-REM}\label{novel_modeling}


In this section, we introduce XT-REM (Extreme-Tail Random Effects Model), which is a novel statistical model that integrates REM and EVT approaches within a unified hierarchical structure. XT-REM employs a threshold to distinguish between standard and extreme proportions, and implements distinct model components in each regime. In this way, it allows for flexible and informative modeling of the distribution of proportions, including rare but clinically important events.

We proceed by presenting the mathematical formulation of the model, its component structure, and the associated likelihood function, which forms the basis for parameter estimation and for establishing its fundamental asymptotic properties, including identifiability and consistency of the maximum likelihood estimators.

\subsection{Model Formulation}


Let $u$ denote the threshold for detecting extreme proportions. We partition the observations into two regimes based on the threshold $u$:

\begin{itemize}
    \item For studies where $p_i \leq u$, the standard REM model is applied as in equations (\ref{theta}) and (\ref{theta_i}).
    
    \item For studies where $p_i > u$, the EVT component is used, the exceedance over the threshold as in equation (\ref{Y})
    is modeled using the GPD.
\end{itemize}

This formulation allows the model to relate the tail characteristics of the proportion distribution to the latent between-study structure. By integrating the REM and GPD components, our novel model simultaneously identifies latent subgroups with elevated risk of extreme proportions, preserves the standard estimation of the aggregate proportion as in the classical REM model, and enhances meta-analytic inference by enabling more precise quantification of tail behavior.

The overall likelihood function for all studies can be written as:
\begin{equation}
\begin{aligned}
&\mathcal{L}
= \prod_{i : p_i \leq u} f_{\mathcal{N}, \mu, \tau^2, \sigma_i^2}(\theta_i)
   \cdot \prod_{i : p_i > u} f_{\text{GPD}, \xi, \beta}(p_i - u) \\
&= \prod_{i : p_i \leq u} \frac{1}{\sqrt{2\pi(\tau^2 + \sigma_i^2)}}
   \exp\left( -\frac{(\theta_i - \mu)^2}{2(\tau^2 + \sigma_i^2)} \right)
   \cdot \prod_{i : p_i > u} \frac{1}{\beta}
   \left( 1 + \frac{\xi}{\beta}(p_i - u) \right)^{-\left( \frac{1}{\xi} + 1 \right)}.
\end{aligned}
\end{equation}

The model partially segments the data into standard and extreme values, allowing for simultaneous modeling of both the central and tail structures of the proportion distribution. This likelihood function describes the probability of the observed set of proportions under the XT-REM model. In a clinical context, it enables simultaneous estimation of the average risk of adverse events and the behavior in the upper tail of the distribution, which corresponds to exceptionally unfavorable outcomes.

The parameters of both distributions are estimated jointly, typically via maximum likelihood, as described in the following section.

\subsection{Parameter Estimation}

After defining the structure of the model, it is necessary to estimate its parameters based on the observed data. The estimated parameters include the fixed effect $\mu$, the variance of the random effects $\tau^2$, as well as the parameters of the GPD component: the baseline shape and scale parameters ($\xi$, $\beta$), and their dependence on the random effects ($\gamma$, $\delta$).

Estimation is conducted using the method of maximum likelihood, which identifies the parameter values that maximize the likelihood of the observed data under the specified model. Due to the hierarchical structure and the combination of different distributions, the model’s log-likelihood is complex and consists of two components, namely the normal REM component and the Pareto EVT component.

We next present the explicit form of the model's log-likelihood function along with the numerical optimization procedure used to obtain parameter estimates. 
We assume that the GPD distribution parameters are identical in all studies for which $p_i > u$, to ensure stable estimation given the typically limited number of exceedances.

For $p_i \leq u$, the REM component is applied. 
The log-likelihood of the REM component is:

\begin{equation}
\ln \mathcal{L}_{\text{REM}} = -\frac{1}{2} \sum_{i \in \mathcal{I}_1} \left[ \ln{(2\pi (\tau^2 + \sigma_i^2))} + \frac{(\theta_i - \mu)^2}{\tau^2 + \sigma_i^2} \right],
\end{equation}
where $\mathcal{I}_1 = \{ i \mid p_i \leq u \}$.

For $p_i > u$, the EVT component is used. 
The EVT log-likelihood is:
\begin{equation}
\ln \mathcal{L}_{\text{EVT}} = -n_2 \ln \beta - \left( \frac{1}{\xi} + 1 \right) \sum_{i \in \mathcal{I}_2} \ln\left(1 + \frac{\xi Y_i}{\beta} \right),
\end{equation}
\noindent where $\mathcal{I}_2 = \{ i \mid p_i > u \}$, and $n_2$ represents the cardinality of $\mathcal{I}_2$.

The total log-likelihood function is the sum of these two components:
\begin{equation}
\ln \mathcal{L} = \ln \mathcal{L}_{\text{REM}} + \ln \mathcal{L}_{\text{EVT}}.
\end{equation}

Numerical maximization of the total log-likelihood can be performed using quasi-Newton optimization algorithms. In particular, the L-BFGS-B method (Limited-memory Broyden–Fletcher–Goldfarb–Shanno with Box constraints) is an efficient variant of the BFGS algorithm, suitable for problems with a large number of parameters and explicit constraints (e.g., $\tau^2 > 0$, $\beta_i > 0$). This method utilizes gradient information and an approximation of the Hessian matrix to efficiently find the optimum while supporting parameter bounds \citep{zhu1997}. When no explicit constraints are imposed, the default procedure used is the standard BFGS algorithm.

We now turn to the choice of the segmentation threshold 
 \( u \), which separates standard from extreme proportions. This threshold can be defined either as a fixed constant or as a dynamic, data-dependent value. In the fixed version, a constant such as \( u = 0.09 \) may be used to facilitate simplicity and direct comparability between different model specifications. Alternatively, a dynamic threshold can be defined based on the latent REM distribution, for example, as its 90th percentile:
\begin{equation}
u = \text{invlogit}(\mu + z_{0.9} \cdot \tau),
\end{equation}
where \( z_{0.9} \approx 1.2816 \), and \textit{invlogit} denotes the inverse logit transformation. The dynamic threshold has the advantage of adapting to the dispersion of the underlying distribution, potentially leading to more stable and context-sensitive model behavior across varying applications.




\subsection{Identifiability of the model}

Identifiability is one of the most important properties of a statistical model, ensuring that
distinct parameter values induce distinct probability distributions for the observed
data. Formally, a parametric model with parameter vector $\Theta$ is identifiable if
\begin{equation}
f(x \mid \Theta_1) = f(x \mid \Theta_2) 
\quad \Longrightarrow \quad \Theta_1 = \Theta_2,
\end{equation}
where $f(x \mid \Theta)$ denotes the joint probability density function of
the observed data \cite{Rothenberg1971}.

In the XT-REM model, identifiability requires particular attention due to the presence of structurally distinct components corresponding to the central and tail parts of the model. These components are separated by a predetermined threshold, which partitions the data into disjoint subsets. Although XT-REM is not a probabilistic mixture model, since no mixing weights or latent allocation variables are introduced, the coexistence of multiple structural components necessitates verification that different parameter values cannot generate the same joint distribution. Classical results on identifiability in mixture and related multi-component models \cite{Teicher1963, Chandra1977, Atienza2006} illustrate the potential difficulties that arise when several sources of variability are modeled simultaneously. In the present setting, however, the deterministic segmentation induced by the fixed threshold 
 simplifies the analysis. In particular, it suffices to verify that no nontrivial reparameterization of the bulk and tail components yields the same joint likelihood. We establish this property under the assumption that the extreme-value component is independent of the random effects and characterized by fixed parameters.


Let $\{p_i\},\ i \in \{1,2,...,N\},$ denote the observed  proportions and let $u \in (0,1)$
be a fixed threshold. The data are partitioned into two disjoint subsets:
$\mathcal{I}_1 = \{ i : p_i \le u \}$ and  $\mathcal{I}_2 = \{ i : p_i > u \}.$

\noindent For studies $i \in \mathcal{I}_1$, the logit-transformed proportions as in the equation (\ref{theta})
are modeled using a random-effects model:
\begin{equation}
\theta_i \sim \mathcal{N}(\mu, \tau^2 + \sigma_i^2),
\end{equation}
where the within-study variances $\sigma_i^2$ are assumed known.

\noindent For studies $i \in \mathcal{I}_2$, the excesses over the threshold as in the equation (\ref{Y})
are modeled using a Generalized Pareto Distribution  with fixed shape ($\xi$) and scale ($\beta$) parameters 
\begin{equation}
Y_i \sim \mathrm{GPD}(\xi, \beta).
\end{equation}

\noindent Therefore, the full parameter vector of the model is:
\begin{equation}
\Theta = (\mu, \tau^2, \xi, \beta).   
\end{equation}

Under the XT-REM model, the joint likelihood factorizes as
\begin{equation}
\mathcal{L}(\Theta) =
\prod_{i \in \mathcal{I}_1}
f_{\mathcal{N}}\!\left( \theta_i \mid \mu, \tau^2 + \sigma_i^2 \right)
\;\cdot \;
\prod_{i \in \mathcal{I}_2}
f_{\mathrm{GPD}}\!\left( Y_i \mid \xi, \beta \right),
\end{equation}
where $f_{\mathcal{N}}$ and $f_{\mathrm{GPD}}$ denote the normal and GPD densities,
respectively.

 Note that the two likelihood components depend on disjoint subsets of observations and on separate subsets of parameters.
The normal random-effects model is identifiable with respect to $(\mu, \tau^2)$ under
standard regularity conditions, provided that the number of studies in $\mathcal{I}_1$ is sufficiently large and the within-study variances exhibit sufficient variability.
Similarly, the GPD is identifiable with respect to
$(\xi, \beta)$ when based on a non-degenerate sample of excesses $\{Y_i\},\ {i \in \mathcal{I}_2}$,
a well-established result in EVT \cite{Coles2001}.

Suppose that two parameter vectors
$
\Theta_1 = (\mu_1, \tau_1^2, \xi_{1}, \beta_{1})$  and\\
$\Theta_2 = (\mu_2, \tau_2^2, \xi_{2}, \beta_{2})$
induce the same joint distribution of the observed data. Then, the equality of the
likelihood contributions from $\mathcal{I}_1$ imply
\begin{equation}
(\mu_1, \tau_1^2) = (\mu_2, \tau_2^2),
\end{equation}
by the identifiability of the normal random-effects component. Likewise, equality of the
likelihood contributions from $\mathcal{I}_2$ imply
\begin{equation}
(\xi_{1}, \beta_{1}) = (\xi_{2}, \beta_{2}),
\end{equation}
by the identifiability of the GPD. Therefore, $\Theta_1 = \Theta_2$, establishing that the XT-REM model is
identifiable.


\subsection{Consistency of the maximum likelihood estimators}

In addition to identifiability, a key asymptotic property of the proposed XT-REM
model is the consistency of the maximum likelihood estimators (MLEs). Consistency
ensures that the estimated parameters converge in probability to their true values
as the number of studies increases. 

Let $\Theta = (\mu, \tau^2, \xi, \beta)$ denote the parameter vector of the 
XT-REM model, and let $\hat{\Theta}_N$ be the corresponding MLE based on $N$ studies.
Under standard regularity conditions for maximum likelihood estimation, consistency
follows from three main requirements: correct model specification, identifiability,
and suitable regularity of the likelihood function \cite{Wald1949, Vaart1998}. 
In hierarchical likelihood settings, these conditions are satisfied provided that the model components are identifiable and the likelihood is well behaved.

We consider an asymptotic framework in which the number of studies $N \to \infty$,
while the individual study sizes remain bounded or stochastic but independent of $N$,
as is standard in meta-analytic settings. The threshold $u$ is assumed fixed and
deterministic. Moreover, we assume that the probability of exceeding the threshold
is strictly positive, i.e., $P(p_i > u) = \pi_u > 0$, so that the number of
exceedances $|\mathcal{I}_2|$ satisfies $|\mathcal{I}_2| \to \infty$ and $\frac{|\mathcal{I}_2|}{N} \to \pi_u$ as
$N \to \infty$. This ensures that both the REM and EVT components are informed
by an increasing number of observations.

In the XT-REM framework, the inclusion of a tail-specific component does not affect the consistency of the central bulk estimators. Since the extreme-value component is fitted to observations exceeding the predetermined threshold, its likelihood contribution remains separated from that of the REM component. This separation principle ensures that standard MLE consistency arguments apply. 

In the XT-REM model, the likelihood function factorizes into two components corresponding to disjoint subsets of the data:
\begin{equation}
\ln \mathcal{L}(\Theta)
=
\ln \mathcal{L}_{\mathrm{REM}}(\mu, \tau^2)
+
\ln \mathcal{L}_{\mathrm{EVT}}(\xi, \beta).
\end{equation}
The REM component corresponds to a standard normal random-effects model for the
logit-transformed proportions and yields consistent MLEs for $(\mu, \tau^2)$ under
classical assumptions. The EVT component is based on the Generalized Pareto
Distribution and produces consistent MLEs for $(\xi, \beta)$ provided that the
number of exceedances above the threshold $u$ increases and the threshold is fixed.

Since the two likelihood components depend on disjoint parameter subsets and are
based on identifiable parametric families, consistency of the full parameter vector
follows directly from the consistency of the individual components, i.e.,
\begin{equation}
\hat{\Theta}_N
=
(\hat{\mu}, \hat{\tau}^2, \hat{\xi}, \hat{\beta})
\;\rightarrow 
(\mu, \tau^2, \xi, \beta), \ \ 
\mbox{as }\ N \to \infty.
\end{equation}
These results establish XT-REM as a well-posed likelihood-based model with sound asymptotic behavior for both its central and tail components.

\section{Simulation Study Design}\label{simulation}
\subsection{Simulation Design}

To evaluate the performance of the proposed XT-REM framework, a Monte Carlo simulation study was conducted. The objective of the experiment was to compare the classical REM with the proposed XT-REM model under varying percentages of extreme studies.

Three simulation scenarios were considered, corresponding to different percentages of extreme studies in the dataset: 5\%, 15\%, and 30\%. These scenarios allow us to assess how model performance changes as the percentage of extreme studies increases. For each Monte Carlo replication ($M = 500$), a dataset consisting of $K = 30$ studies was generated. Study sample sizes $n_i$ were randomly drawn from a uniform distribution on the interval $[100, 1000]$. Central study proportions were generated from a logit-normal random-effects model
\[
\theta_i \sim N(\mu, \tau^2), \qquad p_i = \mathrm{invlogit}(\theta_i),
\]
with parameter values $\mu = \mathrm{logit}(0.035)$ and $\tau = 0.4$, corresponding to a baseline event proportion of approximately $3.5\%$. Extreme proportions were generated using a GPD above a predefined threshold
\[
p_i = u + Y_i, \qquad Y_i \sim \mathrm{GPD}(\xi, \beta),
\]
where $u = 0.09$ denotes the segmentation threshold separating central and extreme observations. 
This threshold was chosen so that it lies well above the typical range of central study proportions 
generated by the logit-normal model, thereby ensuring that the EVT component is applied only to 
unusually large proportions. In particular, the baseline event rate of approximately $3.5\%$ implies 
that proportions exceeding $9\%$ correspond to rare and substantially elevated outcomes. This choice 
provides a clear separation between the central bulk of the data and the extreme tail while ensuring 
that a sufficient number of observations remain available for estimation of the EVT component. 
More generally, the choice of the threshold reflects the standard bias--variance trade-off in the 
peaks-over-threshold approach of Extreme Value Theory: a threshold that is too low may include 
non-extreme observations and introduce bias, whereas a threshold that is too high leads to increased 
variance due to a smaller number of exceedances \cite{Coles2001}. The parameters of the GPD were set 
to $\xi = 0.6$ and $\beta = 0.05$.

Given the true proportions $p_i$ and study sizes $n_i$, the observed number of events was generated according to the binomial model $r_i \sim \mathrm{Bin}(n_i, p_i)$,
and the observed proportions were computed as $\hat{p}_i = \frac{r_i}{n_i}$.

For each simulated dataset, both models were fitted using maximum likelihood estimation. Numerical optimization was performed using the L-BFGS-B algorithm implemented in the \texttt{scipy.optimize} library.

Model performance was evaluated using the following metrics:
\begin{itemize}
\item bias and RMSE of the estimated aggregated proportion,
\item average log-likelihood and AIC,
\item coverage probability of the 95\% confidence interval for the REM estimate,
\item estimated 99\% percentile of the EVT component.
\end{itemize}
All simulations were implemented in Python using the NumPy and SciPy libraries.

\subsection{Simulation Results}

The results of the Monte Carlo simulation study are summarized in Table~\ref{tab:simulation}. 
The three scenarios correspond to increasing percentages of extreme studies in the dataset (5\%, 15\%, and 30\%).


\begin{table}[htbp]
\centering
\caption{Simulation results across three scenarios with increasing percentages of extreme studies.}
\label{tab:simulation}
\begin{tabular}{llcccc}
\toprule
Scenario & Model & Bias & RMSE & AIC & LogLik \\
\midrule
\multirow{2}{*}{S1 (5\%)}  
 & REM     & 0.0041 & 0.0060 & 58.91 & -27.45 \\
 & XT-REM  & \textbf{0.0010} & \textbf{0.0033} & \textbf{36.48} & \textbf{-14.24} \\
\midrule
\multirow{2}{*}{S2 (15\%)} 
 & REM     & 0.0108 & 0.0130 & 75.74 & -35.87 \\
 & XT-REM  & \textbf{0.0013} & \textbf{0.0035} & \textbf{26.67} & \textbf{-9.34} \\
\midrule
\multirow{2}{*}{S3 (30\%)} 
 & REM     & 0.0245 & 0.0267 & 89.09 & -42.55 \\
 & XT-REM  & \textbf{0.0024} & \textbf{0.0043} & \textbf{12.82} & \textbf{-2.41} \\
\bottomrule
\end{tabular}
\end{table}

A clear deterioration in the performance of the classical random-effects model is observed as the percentage of extreme studies increases. In particular, the bias of the REM estimator increases from 0.0041 in Scenario~S1 to 0.0245 in Scenario~S3. A similar pattern is observed for the RMSE, which rises from 0.0060 to 0.0267. In contrast, the proposed XT-REM model remains stable across all scenarios. The bias stays very small, ranging from 0.0010 to 0.0024, while the RMSE increases only slightly from 0.0033 to 0.0043. These results indicate that the XT-REM framework provides more robust estimation when extreme study proportions are present.

\begin{figure}[htbp]
\centering
\includegraphics[width=\textwidth]{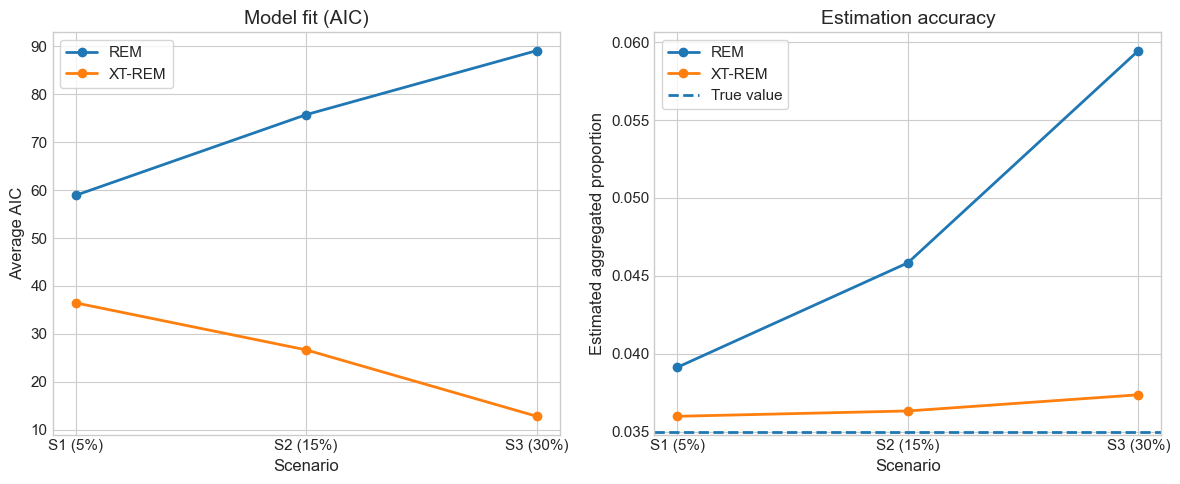}
\caption{Average AIC values and aggregated proportion estimates obtained across simulation scenarios with increasing percentages of extreme studies. The left panel presents the mean AIC values for the REM and XT-REM models, while the right panel shows the estimated aggregated proportions. The dashed horizontal line indicates the true proportion used in the simulation.}
\label{fig:simulation}
\end{figure}

The difference in model fit is further reflected in the log-likelihood and AIC values. As the number of extreme studies increases, the log-likelihood of the REM model decreases considerably, whereas the XT-REM model consistently achieves higher likelihood values. Consequently, the AIC of the classical REM model increases from 58.91 to 89.09 across scenarios, while the XT-REM model achieves substantially lower AIC values, decreasing from 36.48 to 12.82. These trends are illustrated in Figure~\ref{fig:simulation}, where the left panel shows the evolution of AIC values across scenarios, while the right panel presents the estimated aggregated proportions relative to the true value used in the simulation.

\begin{figure}[htbp]
\centering
\includegraphics[width=0.70\textwidth]{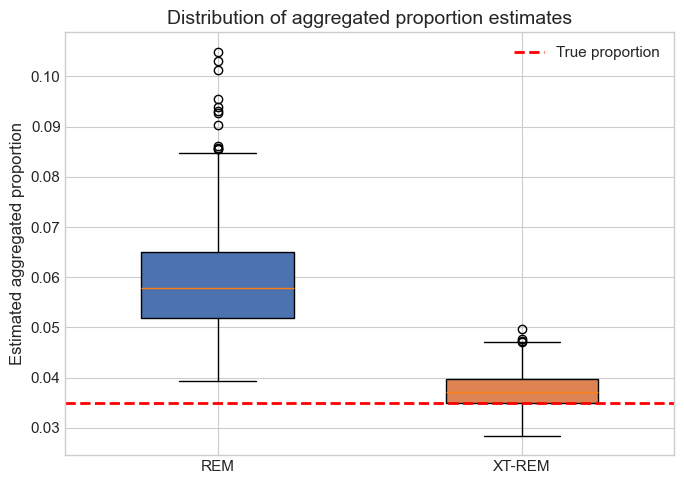}
\caption{Distribution of aggregated proportion estimates across Monte Carlo replications. The dashed horizontal line indicates the true proportion used in the simulation. The XT-REM model produces estimates that are more concentrated around the true value, whereas the classical REM model exhibits larger variability and systematic upward bias.}
\label{fig:distribution}
\end{figure}

To further examine the variability of the estimators across Monte Carlo replications, Figure~\ref{fig:distribution} presents the distribution of aggregated proportion estimates obtained in repeated simulations.

\begin{figure}[htbp]
\centering
\includegraphics[width=0.75\textwidth]{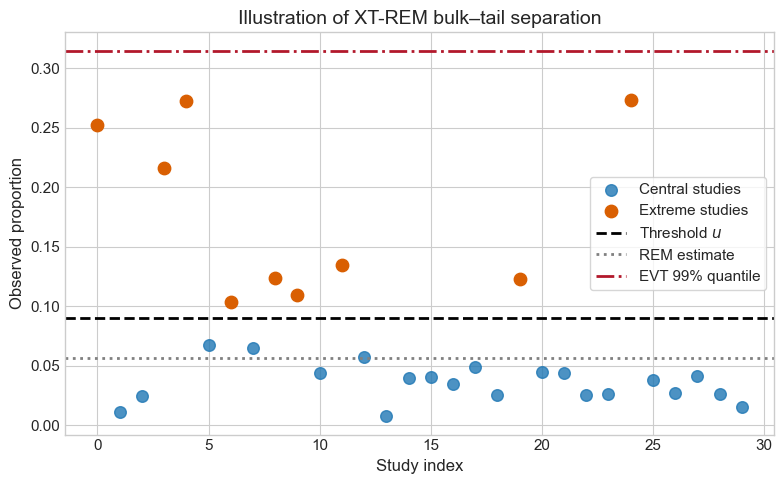}
\caption{Separation of central and extreme observations within the XT-REM framework. Each point represents the observed proportion in a simulated study. Observations below the threshold $u$ are modeled using the REM component, whereas observations exceeding the threshold are treated as extreme values and modeled using the EVT component. The dotted line indicates the REM estimate of the aggregated proportion, and the dash--dot line represents the estimated $99\%$ EVT quantile describing the upper tail behavior.}
\label{fig:xtrem_illustration}
\end{figure}

Finally, Figure~\ref{fig:xtrem_illustration} illustrates the structure of the XT-REM framework by showing how observed study proportions are separated into central and extreme regimes using the threshold $u$.

Taken together, the simulation results indicate that explicitly modeling the extreme tail of the distribution can substantially improve estimation accuracy and model fit when extreme observations are present.

\subsection{Additional Simulation Scenario}

To further examine the behavior of the proposed XT-REM model under a different data-generating mechanism, an additional Monte Carlo simulation scenario was conducted. In this setting, study-level proportions were generated from a logit-normal random-effects model with parameters $\mu = -2.5$ and $\tau = 0.6$, corresponding to a baseline event proportion of approximately $\mathrm{invlogit}(-2.5) \approx 0.076$.

As in the main simulation, extreme observations were introduced using a threshold-based EVT mechanism. A threshold of $u = 0.09$ was used to distinguish between central and extreme observations. With probability $10\%$, study proportions exceeding the threshold were generated from a GPD with parameters $\xi = 0.15$ and $\beta = 0.03$, and then added to the threshold to reconstruct the extreme proportions.
Each simulated dataset consisted of $K = 30$ studies, and the number of events in each study was generated from a binomial distribution with study sample sizes drawn uniformly from the interval $[100,1000]$. A total of $M = 500$ Monte Carlo replications were performed.

Both the classical REM and the XT-REM model were fitted to each simulated dataset using maximum likelihood estimation. Model performance was evaluated using bias and RMSE of the estimated aggregated proportion, as well as the AIC. The results of this additional simulation scenario are summarized in Table~\ref{tab:additional_simulation}.



\begin{table}[htbp]
\centering
\caption{Results of the additional simulation scenario. Bias and RMSE are computed with respect to the true aggregated proportion used in the simulation.}
\label{tab:additional_simulation}

\begin{tabular}{llcccc}
\toprule
Scenario & Model & Bias & RMSE & AIC & LogLik \\
\midrule
\multirow{2}{*}{$S_{ad}$ (10\%)} 
 & REM     & \textbf{0.0065} & \textbf{0.0106} & 58.66 & -27.33 \\
 & XT-REM  & -0.0202 & 0.0208 & \textbf{-32.62} & \textbf{20.31} \\
\bottomrule
\end{tabular}

\end{table}

The results indicate that the classical REM provides slightly more accurate estimation of the central proportion in this scenario, exhibiting smaller bias and RMSE. This behavior is expected since the REM model treats all observations within a single logit-normal framework. In contrast, the XT-REM model explicitly separates extreme observations and models them using the EVT component. While this segmentation introduces a modest bias in the central estimate, the XT-REM model achieves a substantially improved overall model fit, reflected in considerably lower AIC values. This additional scenario highlights the trade-off between accurate estimation of the central parameter and improved modeling of extreme observations. Although the classical REM may yield slightly more accurate estimates of the central effect, explicitly modeling the tail behavior allows the XT-REM framework to provide a more flexible and informative description of the data when extreme proportions are present.


Taken together, the two simulation settings provide a more nuanced picture of the XT-REM framework's performance. In scenarios with a pronounced presence of extreme studies (S1–S3), XT-REM consistently improves both bias and RMSE, while achieving a substantially better likelihood-based fit. However, the additional scenario reveals a trade-off: in strictly homogeneous settings, where the data structure is less aligned with heavy-tail assumptions, the advantages of XT-REM become more selective. While it continues to offer superior AIC and log-likelihood values, the classical REM yields lower bias and RMSE for the central parameter $\mu$. This divergence suggests that the EVT component may occasionally interpret stochastic fluctuations as extreme behavior when the tail is not sufficiently pronounced. Consequently, while XT-REM remains a useful tool for model fit and risk characterization, it is most appropriate in settings where tail behavior is of primary interest or where empirical evidence indicates a departure from normality.
\section{Real Data Study}\label{real_data_study}
\subsection{Data Description}

The real-data analysis is based on a meta-analysis of 16 clinical studies reporting the incidence of pneumonitis in patients treated with immunotherapy. The number of participants per study ranges from 58 to 917, indicating substantial variability in study sizes. The observed proportions of pneumonitis range from approximately 0.6\% to 10.1\%.

Based on the empirical distribution of the observed proportions, a threshold of $u = 0.09$ was selected to separate central and extreme observations. This value lies in the upper tail of the empirical distribution and allows a small number of studies with elevated proportions to be modeled using the EVT component. Studies with proportions exceeding this threshold were classified as extreme and modeled through the EVT part of the XT-REM framework, while the remaining studies were included in the REM component. In total, two studies exceeded the threshold and were assigned to the EVT component, whereas the remaining fourteen studies formed the central REM part. This separation provides a flexible modeling framework in which the central tendency is captured by the random-effects structure, while unusually high proportions are explicitly described using extreme value theory.

\subsection{Estimation of the Combined XT-REM Model}

The combined XT-REM model was estimated using the method of maximum likelihood. The model integrates a classical random-effects meta-analytic component for central observations with an extreme value component for proportions exceeding the previously defined threshold.

For the central observations, the REM component was applied to logit-transformed proportions, assuming a normal distribution with mean $\mu$ and between-study variance $\tau^2$. Within-study variances were treated as known and derived from the binomial sampling model. Proportions exceeding the threshold were modeled through a GPD, characterized by a shape parameter $\xi$ and a scale parameter $\beta$.

The estimated parameters of the XT-REM model are:
\begin{itemize}
    \item REM component: $\hat{\mu} = -3.42$, $\hat{\tau}^2 = 0.36$,
    \item EVT (GPD) component: $\hat{\xi} = -0.14$, $\hat{\beta} = 0.0096$.
\end{itemize}

\begin{figure}[h!]
    \centering
    \includegraphics[width=0.7\textwidth]{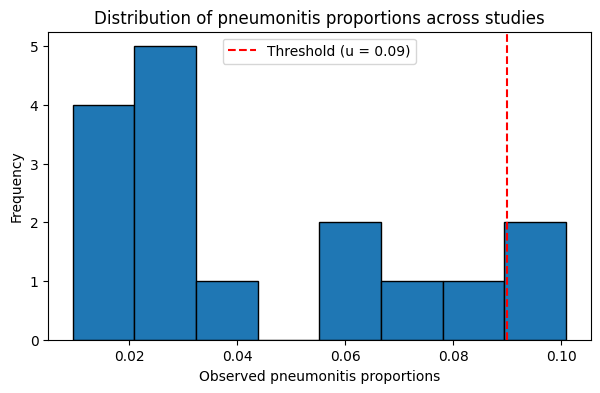}
    \caption{Distribution of observed pneumonitis proportions across studies.}
    \label{fig:pneumonitis_dist}
\end{figure}

Figure~\ref{fig:pneumonitis_dist} illustrates the empirical distribution of pneumonitis proportions across studies. The majority of studies exhibit relatively low incidence rates, while a small number of higher proportions motivate the use of an EVT component to explicitly model the upper tail behavior.

The maximized log-likelihood of the XT-REM model equals 
$\ln L = -6.36$, resulting in
\[
\mathrm{AIC}_{\text{XT-REM}} = 20.73,
\]
with $k=4$ estimated parameters. This represents a substantially improved model fit compared with the classical REM model (AIC = 39.57), despite the increased model complexity.


\subsection{Aggregate and Extreme Proportion Estimates}

The XT-REM model enables a dual interpretation of pneumonitis incidence by 
separately characterizing the central tendency of the data and the behavior of extreme observations. The aggregate estimate is obtained from the REM component, while extreme proportions are modeled through the EVT component. The resulting estimates are summarized in Table~\ref{tab:agg_extreme}.

\begin{table}[h!]
\centering
\caption{Aggregate and extreme pneumonitis proportion estimates based on the XT-REM model.}
\label{tab:agg_extreme}
\begin{tabular}{lcc}
\hline
Component & Estimate & Interpretation \\
\hline
REM (aggregate proportion) & 3.16\% & Central pneumonitis incidence \\
EVT tail (99th percentile) & 12.3\% & Extreme incidence level \\
\hline
\end{tabular}
\end{table}

The REM component yields an overall pneumonitis incidence estimate of 
approximately 3.2\%, representing the central tendency across studies. 
The EVT component captures the upper-tail behavior of the distribution, 
suggesting that extreme pneumonitis proportions may reach values slightly 
above 12\%. Although only a small number of studies exceed the predefined 
threshold, modeling these observations separately improves the robustness 
of the overall inference.

\begin{figure}[h!]
    \centering
    \includegraphics[width=0.7\textwidth]{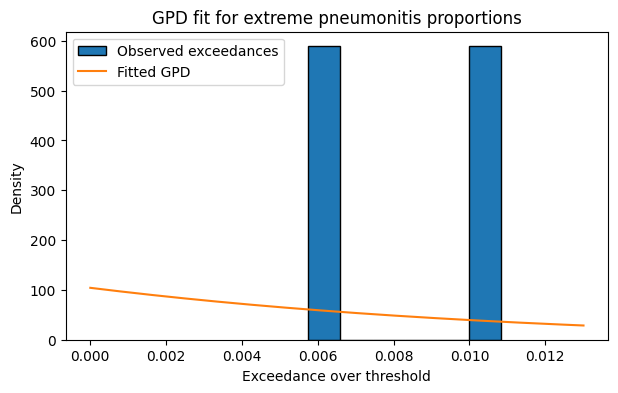}
    \caption {Fitted GPD for exceedances of pneumonitis proportions above the selected threshold. The histogram shows observed exceedances and the curve represents the fitted GPD density.
}
    \label{fig:gpd_fit}
\end{figure}

Figure~\ref{fig:gpd_fit} illustrates the fitted GPD for exceedances above the selected threshold. Only a small number of studies exceed the threshold, resulting in limited data for tail modeling. Nevertheless, the fitted GPD suggests that extreme pneumonitis proportions may reach values slightly above 10\%, supporting the inclusion of an EVT component while highlighting the exploratory nature of the tail estimation.

\subsection{Comparison with the Classical REM Model}

For comparison, the classical REM was fitted to the complete dataset without separating extreme observations. The estimated parameters were
\[
\hat{\mu}_{\mathrm{REM}} = -3.25, \quad
\hat{\tau}^2_{\mathrm{REM}} = 0.37,
\]
with maximized log-likelihood $\ln L_{\mathrm{REM}}=-17.78$ and
\[
\mathrm{AIC}_{\mathrm{REM}} = 39.57.
\]

The corresponding aggregate pneumonitis proportion estimate was approximately 3.74\%. In contrast, the XT-REM model achieved a substantially lower AIC value (20.73), indicating improved model fit despite the additional EVT parameters. This result suggests that explicitly accounting for extreme observations leads to a more stable and realistic characterization of pneumonitis incidence in meta-analytic settings.


In this real-data analysis, the XT-REM model provides several advantages over the classical REM approach when analyzing meta-analytic data with potential extreme proportions. First, the XT-REM model yields a slightly lower aggregate estimate of the central pneumonitis incidence (approximately 3.2\% compared with 3.7\% under the classical REM model). This difference arises because extreme observations are modeled separately through the EVT component, which leads to a reduced influence of these observations on the estimation of the central random-effects structure. Second, the EVT component enables an explicit, though data-limited, characterization of the upper tail of the distribution. While the REM model implicitly assumes a symmetric logit-normal structure, the XT-REM framework provides quantitative insight into extreme proportions. In the present analysis, the estimated 99th percentile suggests that extreme pneumonitis incidences may reach values slightly above 12\%, although this estimate should be interpreted cautiously due to the limited number of extreme observations. Third, the combined model achieves a substantially improved statistical fit. The XT-REM model attains a considerably lower Akaike Information Criterion (AIC = 20.73) compared with the classical REM model (AIC = 39.57), indicating a better fit for the observed data within this modeling framework. Finally, unlike approaches that treat extreme observations as outliers to be removed, the XT-REM model explicitly incorporates them into the analysis. This preserves information contained in the tail of the distribution while maintaining a stable estimation of the central tendency.

Overall, the XT-REM framework provides a flexible and informative extension of classical random-effects meta-analysis, allowing simultaneous modeling of typical study outcomes and rare but clinically relevant extreme events, while requiring cautious interpretation when the number of extreme observations is limited.

\section{Conclusion}\label{conclusion}

In this study, the XT-REM model is introduced as a two-component framework for the meta-analysis of proportions, integrating the standard random-effects paradigm with extreme value theory. The model enables partial segmentation of the included studies based on a pre-specified threshold and supports independent modeling of the central and tail components of the proportion distribution. This approach is particularly relevant in clinical settings where high proportions of adverse events are infrequent yet clinically significant.

Empirical validation based on data from a meta-analysis of immunotherapy-related adverse events is consistent with the overall findings of the simulation study. The XT-REM model demonstrates improved likelihood-based fit compared with the conventional random-effects model and enables a more explicit characterization of extreme proportions. Simulation experiments further indicate that XT-REM can reduce bias and root mean squared error in settings with a pronounced presence of extreme observations, while in more homogeneous scenarios its advantages are less pronounced and primarily related to model fit.

Future research directions include a systematic assessment of the model’s sensitivity to threshold selection, as well as the exploration of alternative specifications of the EVT component. In particular, a promising extension involves introducing a hierarchical dependence structure in which the parameters of the GPD distribution are allowed to depend on the latent random effects from the REM component. Such an approach would establish a direct link between between-study heterogeneity and tail behavior, potentially leading to a more coherent joint modeling framework, but it also raises additional theoretical and computational challenges. Further developments may also include an extension of the framework toward dynamic models with time-varying effects or the incorporation of covariate information. 

The model’s interpretability and flexibility suggest that it may be useful across diverse clinical and epidemiological domains, where extreme values are not only expected but also critical for informed decision-making.

\section*{Acknowledgements}\label{acknowledgements}
This research was funded by the Science Fund of the Republic of Serbia (Grant No. 9393) through the project "Optimization and Prediction in Therapy Treatments of Cancer – OPTIC", and by the Ministry of Science, Technological Development and Innovation (Contract No. 451-03-34/2026-03/200156) and the Faculty of Technical Sciences, University of Novi Sad through project “Scientific and Artistic Research Work of Researchers in Teaching and Associate Positions at the Faculty of Technical Sciences, University of Novi Sad 2026” (No. 01-3609/1).














\end{document}